\documentclass{article}
    \PassOptionsToPackage{numbers, compress}{natbib}
\usepackage[dblblindworkshop, preprint]{neurips_2025}

\usepackage[utf8]{inputenc} 
\usepackage[T1]{fontenc}    
\usepackage{hyperref}       
\usepackage{url}            
\usepackage{booktabs}       
\usepackage{amsfonts}       
\usepackage{nicefrac}       
\usepackage{microtype}      
\usepackage{xcolor}         
\usepackage{amsmath}
\usepackage{graphicx}
\usepackage{float}
\usepackage{booktabs}
\usepackage{caption}
\usepackage{enumitem}
\usepackage{hyperref}
\usepackage{fvextra}
\usepackage{minted}
\usepackage{arydshln} 
\usepackage{threeparttable}
\usepackage{tablefootnote}
\usepackage{graphicx}
\usepackage{subcaption}
\usepackage{wrapfig}
\usepackage{tcolorbox}
\usepackage{amsthm}
\usepackage{enumitem}

\tcbset{
  blue/.style={
    colback=blue!5!white,    
    colframe=blue!40!black,  
    boxrule=0.6pt,           
    arc=2mm,                 
    left=4pt, right=4pt, top=4pt, bottom=4pt, 
  }
}

\title{\name: Safeguarding Privacy for LLM-Based \Ts\ via Abstraction}

\author{
  Sepideh Abedini$^{1,2}$ \quad
  Shubhankar Mohapatra$^{1}$ \quad
  D. B. Emerson$^{2}$ \quad \\
  \textbf{Masoumeh Shafieinejad$^{2}$} \quad
  \textbf{Jesse C. Cresswell$^{3}$} \quad
  \textbf{Xi He$^{1,2}$} \\
  University of Waterloo$^{1}$ \\
  \texttt{\{sepideh.abedini,shubhankar.mohapatra,xi.he\}@uwaterloo.ca} \\
    \begin{tabular}{cc}
      Vector Institute$^{2}$ & Layer 6 AI$^{3}$ \\
      \texttt{\{david.emerson, masoumeh\}@vectorinstitute.ai} &
  \texttt{jesse@layer6.ai}
    \end{tabular}
} 
\makeatletter
\renewcommand{\@noticestring}{}
\makeatother

\begin{document}
\newcommand{\name}{MaskSQL}
\newcommand{\ts}{text-to-SQL}
\newcommand{\Ts}{Text-to-SQL}
\newcommand{\signpost}[1]{\noindent\textbf{#1.}}
\newcommand{\red}[1]{{\color{red}{#1}}}
\newcommand{\shub}[1]{{\color{blue}{Shub: #1}}}
\newcommand{\Mas}[1]{{\color{purple} (Mas: #1)}}
\definecolor{dgreen}{HTML}{228B22}
\newcommand{\sepid}[1]{{\color{dgreen}(Sepideh: #1)}}
\newcommand{\dbe}[1]{{\color{orange} (DBE: #1)}}

\newcommand{\schema}{\ensuremath{\mathcal{S}}}
\newcommand{\ques}{\ensuremath{\mathcal{Q}}}
\newcommand{\prompt}{\ensuremath{\mathcal{P}}}
\newcommand{\policy}{\Psi}
\newcommand{\tables}{\ensuremath{\mathcal{T}}}
\newcommand{\columns}{\ensuremath{\mathcal{C}}}
\newcommand{\values}{\ensuremath{\mathcal{V}}}
\newcommand{\sql}{\ensuremath{\mathcal{Y}}}
\newcommand{\constraints}{\ensuremath{\Phi}}
\newcommand{\bird}{BIRD}
\newcommand{\din}{DIN-SQL}
\newcommand{\dail}{DAIL-SQL}
\newcommand{\qwen}{Qwen-2.5-7B-Instruct}
\newcommand{\gpt}{GPT-4.1}

\newtheorem{defn}{Definition}[section]
\newtheorem{exa}[defn]{Example}
\newtheorem{example}{Example}

\newcommand{\squishlist}{
	\begin{list}{$\bullet$}
		{
			\setlength{\itemsep}{0pt}
			\setlength{\parsep}{0pt}
			\setlength{\topsep}{3pt}
			\setlength{\partopsep}{0pt}
			\setlength{\leftmargin}{1.5em}
			\setlength{\labelwidth}{1em}
			\setlength{\labelsep}{0.5em} } }

	\newcommand{\squishend}{
\end{list}  }

\newlength{\itemizelength} 
\setlength{\itemizelength}{-20pt}

\maketitle

\begin{abstract}
Large language models (LLMs) have shown promising performance on tasks that require reasoning, such as \ts\ translation, code generation, and debugging. However, regulatory frameworks with strict privacy requirements constrain their integration into sensitive systems. State-of-the-art LLMs are also proprietary, costly, and resource-intensive, making local deployment impractical. Consequently, utilizing such LLMs often requires sharing data with third-party providers, raising privacy concerns and risking noncompliance with regulations. Although fine-tuned small language models (SLMs) can outperform LLMs on certain tasks and be deployed locally to mitigate privacy concerns, they underperform on more complex tasks such as \ts\ translation. 
In this work, we introduce \name, a \ts\ framework that utilizes abstraction as a privacy protection mechanism to mask sensitive information in LLM prompts. Unlike redaction, which removes content entirely, or generalization, which broadens tokens, abstraction retains essential information while discarding unnecessary details, striking an effective privacy–utility balance for the \ts\ task. Moreover, by providing mechanisms to control the privacy-utility tradeoff, \name\ facilitates adoption across a broader range of use cases.
Our experimental results show that \name\ outperforms leading SLM-based \ts\ models and achieves performance approaching state-of-the-art LLM-based models, while preserving privacy. Our code is available at \href{https://github.com/sepideh-abedini/MaskSQL}{\texttt{https://github.com/sepideh-abedini/MaskSQL}}.
\end{abstract}

\section{Introduction}\label{sec:intro}
Structured databases are central to applications across science, business, healthcare, and government. However, retrieving information from these systems typically requires knowledge of Structured Query Language (SQL), creating a steep barrier for non-technical users. The \ts\ task bridges this gap by translating natural language (NL) questions into executable SQL queries, allowing users to interact with databases intuitively and without specialized expertise.

Recent advances in language models (LMs) have significantly improved the accuracy and availability of \ts~solutions, enabling deployment across diverse domains and schemas. For example, the most performant approaches on the \ts~benchmarks Spider \cite{spider},
\bird \citep{bird}, and Spider 2.0 \cite{spider2} rely on LMs as their backbone for SQL generation. An LM's performance typically scales with parameter count \citep{slm_tradeoff}, leading to a common distinction \cite{slm-llm-size} between small LMs (SLMs, $<10$B parameters), and large LMs (LLMs, 10s to 100s of billions of parameters) which require specialized infrastructure to run. While LLMs dominate \ts~benchmarks, due to their increased hardware requirements they are typically accessed via remote APIs hosted by specialized inference providers. These hosted APIs are attractive to users as they largely eliminate infrastructure outlays, but also introduce severe privacy concerns; passing data through third-party APIs exposes sensitive user and schema information to heightened privacy risks \citep{rel:llm-leak-25, rel:llm-ban-23, LLM-limitations}. For instance, recent work demonstrated that databases used in a \ts\ system are vulnerable to schema inference, where an adversary can reconstruct proprietary schema details by probing the system with carefully crafted queries \citep{inference_attack}. These criticisms could lead to stricter standards for \ts\ tasks in privacy laws such as GDPR in Europe, HIPAA in the USA, or PIPEDA in Canada.


\begin{wrapfigure}[24]{r}{0.5\textwidth}
    \vspace{-18pt}
    \begin{tcolorbox}[blue, width=0.48\textwidth]
        \begin{example}\label{running_example}
Assume a database schema with three tables that contain sensitive 
information about patients in hospitals in New York:
\begin{align*}
T_1 &: \text{Patients}(pid, name, hiv\_status, \\& \qquad diagnosis, treatment) \\
T_2 &: \text{Hospital}(hid, name, address) \\
T_3 &: \text{Admissions}(aid, pid, hid, date)
\end{align*}
A doctor with minimal database experience wants to generate an SQL query for the question $\ques$: ``How many patients did the New York Hospital admit with HIV status as positive?''. The doctor sends this query along with the database schema $\schema$ to an LLM hosted remotely to generate the corresponding SQL query $\sql$. Although the LLM provider lacks access to the database and cannot execute the generated query $\sql$, the prompt has disclosed sensitive information: the existence of a table named ``Patients'', a column named ``hiv\_status'', and possible literal values, such as ``positive''. 
\end{example}
    \end{tcolorbox}
\end{wrapfigure}

To illustrate the dilemma, consider the \ts\ task shown in Example \ref{running_example}. The user has two options to generate this query: (i) send their data to a powerful but \emph{untrusted} LLM hosted remotely, with the risk of exposing schema and personally identifiable information (PII); or (ii) rely on \emph{trusted} SLMs hosted locally, which often fail to handle complex SQL constructs such as nested queries, window functions, or common table expressions (see Figure \ref{fig:hard-sql-example} in Appendix \ref{sec:extra-figs}). Notable downsides exist with either option.

In this work, we propose a third alternative: \name, a privacy-preserving \ts\ framework that combines the utility of LLMs with the trust guarantees of local processing using SLMs. \name\ achieves this through \emph{prompt abstraction}, which systematically replaces sensitive schema elements (table names, column names, and cell values) with abstract symbols before sending a \ts\ prompt to a remote LLM. Upon receiving the LLM's output, the SQL query is reconstructed locally to restore the abstracted values to their original values and make the query valid and executable.

However, implementing abstraction for \ts\ presents unique challenges: (i)
accurately identifying sensitive tokens in the NL question according to user-defined privacy policies; (ii) preserving the utility of both the NL question and database schema in the abstraction process for accurate SQL generation; and (iii) correcting errors introduced by abstraction noise. \name\ addresses these with a three-stage pipeline of abstraction, SQL generation, and SQL reconstruction. On a challenging subset of the BIRD benchmark, \name\ outperforms state-of-the-art SLM-based approaches in accuracy, while preserving the privacy of user input, unlike LLM-based approaches. 

The main contributions of this work are as follows.
\begin{enumerate}
\setlength{\leftskip}{\itemizelength}
    \item We formalize the problem of privacy-preserving \ts\ using prompt abstraction guided by user-defined privacy policies.
    \item We introduce \name, a framework that safeguards privacy when using untrusted remote LLMs by abstracting sensitive information in the question and database schema.
    \item We propose a policy-based abstraction mechanism that enables a privacy–utility tradeoff, allowing users to customize the level of abstraction according to their needs.
    \item We empirically evaluate \name\ on $300$ complex queries from the BIRD benchmark, showing that it surpasses trusted SLM-based methods in accuracy while mitigating schema and PII leakage.
\end{enumerate}
\section{Background}

In this section, we follow \cite{benchmarking} to formalize the \ts\ task and provide some requisite background on 
LMs. We define a database schema as a tuple $\schema = (\tables, \columns, \constraints)$ where $\tables = \{T_1, T_2,\ldots\}$ is the 
set of tables, $\columns = \{C_1,C_2,\ldots\}$ is the set of columns across all tables, and $\constraints = \{\phi_1, \phi_2, \ldots\}$ is the set of database constraints that enforce links between tables, such as primary or foreign keys. Each $C_i$ is also associated with a set of literal values, $V_i$, present in the database such that $V_i \subset \text{dom}(C_i)$, where $\text{dom}(C_i)$ is the domain of column $C_i$. For example, $V_i$ could represent real numbers for a numerical column, or textual categories for a categorical column. Herein, we consider access only to the schema $\schema$ and not the values $V_i \in \values$, where $\values$ is the set of all literal values in the database.


\signpost{\Ts\ Task Formulation} 
\Ts~translation aims to map an 
NL question to its corresponding executable 
SQL code with respect to a given database schema \citep{slm_survey}. Formally, given a natural language question $\ques$ and a database schema $\schema$, the task is to generate a SQL query $\sql$, such that $\sql$ is executable and accurately represents the intent of $\ques$. $\ques$ is represented as a sequence of tokens such that $\ques = w_1,\ldots,w_n$ for $w_i \in \mathbb{W}$, where $\mathbb{W}$ is the vocabulary of permissible tokens (we use the English language vocabulary).


\signpost{Language Models}
Language Models are neural models trained on large-scale corpora of natural language and structured text. They are often informally classified into two types based on the total number of parameters. LLMs have a very high parameter count (tens to hundreds of billions) which requires multiple GPUs or specialized hardware to run, and, hence, are usually accessed through a third-party service. In contrast, SLMs use fewer parameters (less than ten billion) and can typically be run on a single GPU, enabling them to be hosted locally. Due to the difference in model complexity, SLMs have reduced utility and ability to reason compared with LLMs \cite{emergent}.

LMs are used by providing a prompt 
to specify the task to be performed. Hosted LLMs require the prompt to be passed through an API to the remote infrastructure. Once this happens, the owner loses control over how their data is processed, used, or stored, which is a major concern when there is an obligation to protect that data. Compliance with relevant laws and regulations around data privacy may preclude the use of hosted LLMs due to this loss of control. Thus, there is a need for solutions that balance the performance of LLMs with the increased control and privacy that SLMs bring.

In the context of this paper, we use LMs for the generation of SQL queries from NL questions. Given question $\ques$ and schema $\schema$, we form a prompt $P(\ques, \schema)$ which is input to an LM denoted by $f_\text{LM}$. The \ts\ task is represented as $\sql = f_{\text{LM}}(P(\ques,\schema))$, where $\sql$ is the generated SQL query.


\section{Related Work}\label{sec:related}

Text-to-SQL translation has been extensively studied in both the NLP and database communities. Earlier methods relied on rule-based systems and named-entity recognition (NER)~\citep{Baik2020DuoquestAD, Quamar2022NaturalLI}, followed by neural approaches using LSTMs~\citep{wang2019rat, seq2sql, sqlnet, yu2018typesql} and transformers~\citep{lei2020re, ma2020mention}. More recently, prompt-based LLMs have emerged as the state-of-the-art~\citep{din,dail}. Comprehensive comparisons of these techniques are provided on benchmarks including Spider \cite{spider}, BIRD~\citep{bird}, Spider 2.0~\cite{spider2}, and in recent surveys~\citep{zhu2024large, tssurvey, deng2022recent}.
An alternative line of work focuses on locally served SLMs. Several methods~\citep{dts,codes,msc,slm-sql} improve SLM performance via fine-tuning. While using local SLMs eliminates the privacy concerns of sharing data with third parties, they consistently underperform on complex queries requiring stronger reasoning~\citep{slm-sql}. Our approach follows a hybrid strategy, combining SLMs and LLMs, and we evaluate it against some state-of-the-art models reported on the \bird~leaderboard.\footnote{\hyperlink{\bird~ Leaderboard}{https://bird-bench.github.io/}}

A parallel line of work emphasizes privacy-preserving LLM inference, as user prompts often contain sensitive information. Cryptographic approaches include homomorphic encryption (HE)~\citep{rel:llm-he} and multi-party computation (MPC)~\citep{rel:llm-mpc}, which protect user data during inference but incur significant computational overhead. Differential privacy (DP) has also been applied to preserve training privacy~\citep{rel:dp-llm-1, rel:dp-bart, rel:dp-pre, rel:dp-learn} or inference privacy using noisy embeddings~\citep{rel:dp-llm-denoise, rel:dp-llm-fw} and token substitutions~\citep{rel:infer-dpt, rel:santext}.
However, such methods either do not operate well at the scale of LLMs, or degrade model utility~\citep{rel:survey}—a critical limitation for \ts, where precise information retrieval is required. In contrast, we propose a practical privacy protection approach, leveraging local SLMs and prompt abstraction to minimize the exposure of sensitive data while preserving semantic fidelity.

Recent work has explored prompt sanitization for LLM inference privacy~\citep{rel:has, rel:port, rel:papillon, rel:san, rel:preempt}.
For example, Portcullis~\citep{rel:port} employs NER to detect sensitive entities, while PP-TS~\citep{rel:san} and Pr$\varepsilon\varepsilon$mpt~\citep{rel:preempt} sanitize inputs by replacing PII with contextually appropriate surrogates. HaS~\citep{rel:has} and Papillon~\citep{rel:papillon} rely on fine-tuned local models for anonymization. Related to these are generalization-based methods~\cite{general, rel:incognito, rel:optimal-k}, which replace values with general terms to obscure specifics. 
Unlike these general-purpose techniques, the abstraction process we apply for \ts\ 
consistently preserves the alignment between the question's logic and the database schema.

\section{Problem Statement}\label{sec:prob}

Consider a user with limited local compute resources who, given their database schema $\schema$ and an NL question $\ques$, wants to generate a SQL query $\sql$ that performs the task described in $\ques$. The user could be, for example, a financial institution or a hospital that holds sensitive information about individuals and is prohibited from sharing that information with untrusted third parties. As such, we assume that any remotely hosted LLM is untrusted \cite{privacy-survey}. 
Specifically, we consider a privacy policy $\policy$ that defines what information in $\schema$, $\ques$, or the database values $\values$ is considered sensitive. A privacy policy is determined by the user according to their needs and may contain all or a subset of the following information.
\begin{itemize}
\setlength{\leftskip}{\itemizelength}
\item \textbf{Table and Column Names:} Any table or columns names from the sets of tables $\tables$ and columns $\columns$ defined in $\schema$, including words or terms in
$\ques$ that refer such tables or columns.
\item \textbf{Literal Values:} Any words or terms in $\ques$ 
that might refer to any cell value from the set $\values$ in the database. 
\end{itemize}
By default, we use the full policy $\policy_{F}$ which includes the entire schema $\schema$ and all values $\values$, as sensitive. However, less restrictive policies can also be defined, depending on the user's requirements. 
The problem of privacy-preserving \ts\ is to leverage a LM to accurately generate an SQL query $\sql$ that correctly implements the intent of $\ques$ without exposing any information defined in the policy $\policy$.

\section{\name: Privacy-Preserving \Ts\ Generation}

Our proposed approach, \name, leverages LLM capabilities while protecting sensitive information. We identify that, for a model to generate a correct SQL query $\sql$, the essential information in the prompt is the mapping between the terms used in the NL question $\ques$ and entities in the database schema $\schema$. Specific names or values are not crucial for the structure and syntax of the generated query. Thus, table names, column names, and literal values can be abstracted away in the prompt and later restored by using a bijective mapping between the original and abstract tokens. Consider Example \ref{running_example}: \ques:  ``\emph{How many patients did the New York Hospital admit with HIV status as positive?}''. 
 An abstracted query, ``\emph{How many $T_1$ did the $V_1$ $T_3$ with $C_3$ as $V_2$?''}, paired with an appropriately abstracted schema retains all information needed to generate $\sql$. 
 
There are multiple challenges with respect to implementing such an abstraction approach in practice. First, the tokens in $\ques$ must be accurately linked to the corresponding elements in $\schema$. This is difficult because a non-technical user may enter tokens in $\ques$ that do not exactly match those used in $\schema$. For instance, the user may write the token \emph{admit} which must be linked to the table named \emph{Admissions} based on the semantic context. Second, the abstracted prompt is processed by an LLM, which generates a similarly abstracted SQL query. The abstracted SQL query then needs to be reliably mapped back to its concrete form. This step needs to be performed accurately, as incorrect mapping can lead to syntax errors (e.g. joins applied with incorrect column names). Third, $\sql$ may contain minor errors or issues (e.g., using the string value of \emph{positive} instead of the numeric value of $1$) that must be corrected. These errors can occur due to noisy translation during the process and must be corrected for error-free query execution. 

\begin{figure}[t]
    \centering
    \includegraphics[width=\linewidth]{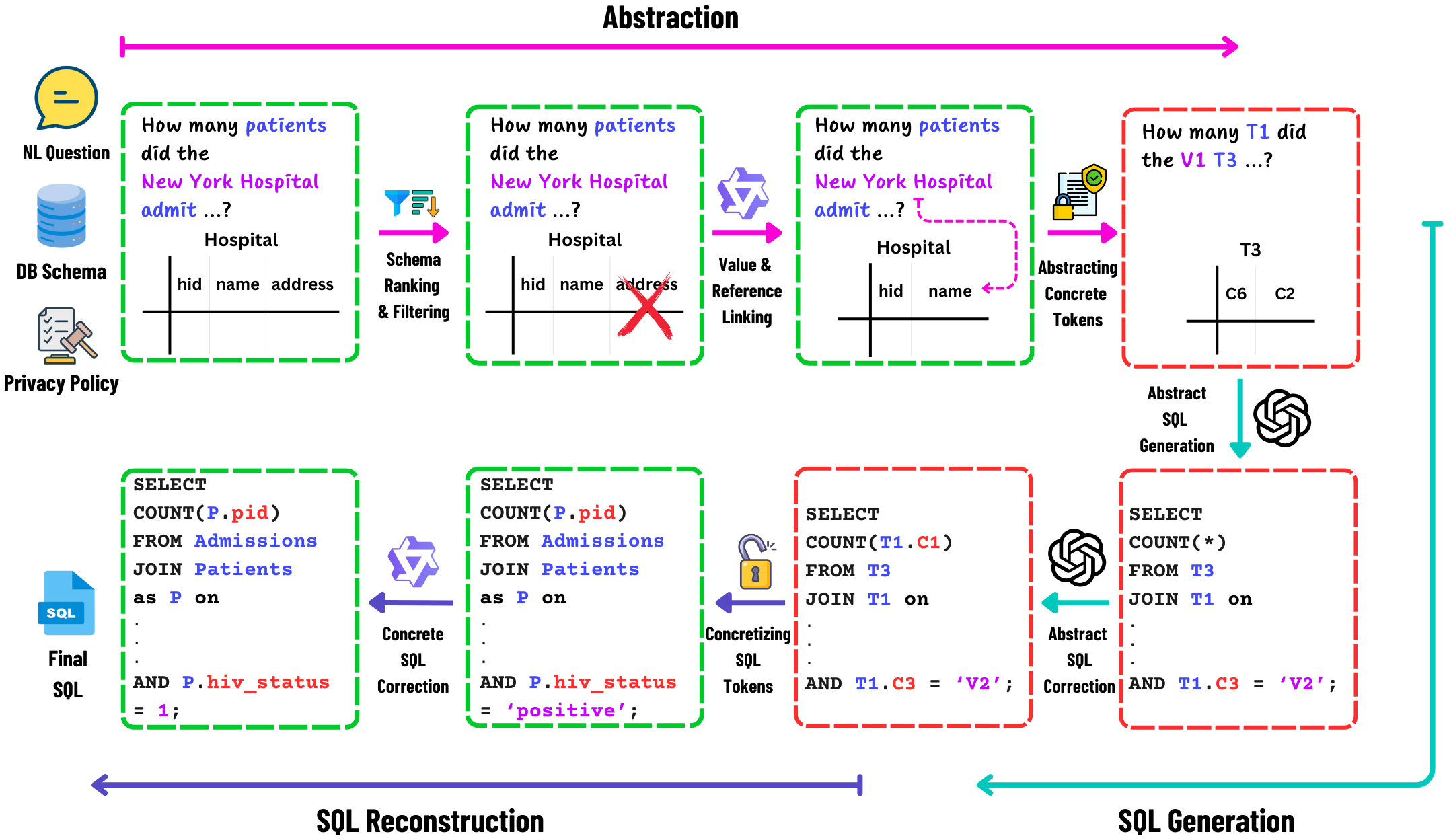}
    \caption{\name\ pipeline. Green dashed boxes delineate text and schema information contained in the \textit{``trusted environment''}, while red boxes denote those exposed to \textit{``untrusted third parties''}.}
    \label{fig:pipe}
\end{figure}

\name\ addresses these challenges within a three-stage pipeline: \emph{Abstraction}, \emph{SQL Generation}, and \emph{SQL Reconstruction}. 
Altogether, this pipeline safeguards user privacy while leveraging the capabilities of LLMs. An overview of the pipeline is shown in Figure \ref{fig:pipe}. In the first stage, $\ques$, $\schema$, and $\policy$ are passed to the \emph{Abstraction} engine which produces abstracted representations $\ques'$ and $\schema'$. In the second stage, these abstracted inputs are combined into a prompt $P(\ques', \schema')$ and provided to the \emph{SQL Generation} engine, which prompts a remotely hosted LLM and returns an abstracted SQL query $\sql'$. In the final stage, the \emph{SQL Reconstruction} engine maps $\sql'$ back to its concrete form $\sql$, applies a final round of corrections to fix any remaining errors, and outputs the executable SQL query. Figure \ref{fig:running-example-abs} in Appendix \ref{sec:extra-figs} shows $\ques'$, $\schema'$, and $\sql'$ under the full policy, $\policy_{F}$, for Example \ref{running_example}. The remainder of this section describes each stage in detail.

\subsection{Abstraction}
The first stage of the pipeline generates abstract versions of both $\ques$ and $\schema$. This process consists of three main steps.
\begin{enumerate}
\setlength{\leftskip}{\itemizelength}
    \item Ranking and filtering entire elements of $\schema$ to drop irrelevant tables and columns based on $\ques$.
    \item Identifying mappings between sensitive terms in $\ques$ and the retained schema elements using a locally hosted SLM.
    \item Replacing the identified terms in $\ques$ and the retained schema elements with abstract identifiers.
\end{enumerate}
The 
bijective masking procedure, combined with SLM-based schema linking, provides an effective approach for abstracting a \ts~query. We detail each of these steps below.


\signpost{Schema Ranking and Filtering} 
Real-world databases often include hundreds of tables and columns, which can overwhelm LMs with irrelevant context. To mitigate this, we use a local cross-encoder model, following the methodology of RESDSQL \citep{resdsql}, to rank and filter schema elements based on their relevance to $\ques$. Specifically, RESDSQL employs a RoBERTa-based cross-encoder \cite{roberta} to compute contextual embeddings for both $\ques$ and $\schema$. These embeddings are pooled using a Bidirectional Long Short-Term Memory \cite{bilstm} and scored via Multi-Layer Perceptron modules to estimate the relevance of each table and column. Based on these scores, the top-$k$ tables and their corresponding top-$j$ columns are retained. This strategy is integrated into the pipeline, as keeping too few elements may result in missing relevant matches, while too many can introduce noise and increase computational overhead. As shown in Appendix \ref{sec:abl}, schema filtering has a significant impact on both accuracy and efficiency. In our experiments, retaining $k=4$ tables and $j=5$ columns per table yields strong performance, though these parameters can be adjusted based on specific use cases. The output of this step is a ranked and filtered list of relevant schema elements, which are then passed to subsequent stages.

\signpost{Value and Reference Linking}
This step takes as input the filtered list of schema elements from the previous step and $\ques$, and constructs a mapping between the two using a local SLM. This process is performed in three sub-steps. First, the SLM is prompted to identify tokens $w_i \in \ques$ that may correspond to values in the database, $\values$. In Example \ref{running_example}, the tokens \textit{New York Hospital} and \textit{positive} are identified. In the second step, the SLM is prompted to map the identified values to their corresponding columns and tables, from $\columns$ and $\tables$ in $\schema$, respectively. For Example \ref{running_example}, \textit{New York Hospital} is mapped to \textit{Hospital.name} and \textit{positive} is mapped to \textit{Patients.hiv\_status}. In the final step, the SLM is prompted to identify any remaining tokens $w_i \in \ques$ that may reference any column or table names in the filtered schema elements. In Example \ref{running_example}, tokens \textit{patients}, \textit{admit}, and \textit{HIV status} are mapped to the tables \textit{Patients} and \textit{Admissions}, and column \textit{Patients.hiv\_status}, respectively. These mappings are then passed to the next step.

The importance of value and reference linking is two-fold. To protect privacy and abstract away sensitive tokens, it is essential to generate an accurate and complete linking; any token not identified as a value or reference remains unmasked and, therefore, may be exposed. In addition, the linking map must be generated precisely to preserve reference information during the abstraction process. 
Also, prior work shows that SLMs achieve accuracy comparable to LLMs for schema linking tasks \citep{death}, making them sufficient for this stage.


\signpost{Abstracting Concrete Tokens} 
\label{sec:trade-off}
The final step generates the abstracted NL question $\ques'$ and the schema $\schema'$. To generate $\schema'$, each table, column, and value specified in $\policy$ and retained after the initial schema filtering step is assigned an abstract symbol. For example, table names are mapped to symbols like $T_i$, and columns to $C_i$. Tokens that are not included in the privacy policy $\policy$ remain unchanged. The mapping from $\schema$ to $\schema'$ is stored as a symbol lookup table, which is later used for both masking and reconstruction. Next, using this symbol table and the linking map from the previous step, all references to tables and columns in $\ques$ are replaced with their corresponding abstract symbols. In Example \ref{running_example}, the table name \textit{Patients} is represented by the symbol $T_1$, column \textit{Patients.pid} by symbol $C_1$, and column \textit{Patients.name} by symbol $C_2$, and so on. 

Finally, each literal value in $\ques$ is also replaced with a unique abstract symbol of the form $V_i$. As shown in Figure \ref{fig:pipe}, the token \textit{New York Hospital} is mapped to $V_1$. An additional sentence is also appended to the question to specify the column associated with the value. In Example, \ref{running_example}, the sentence: ``$V_1$ is a value of the column $C_7$.'' where $C_7$ is the abstracted symbol for column \textit{Hospital.name}, is appended to the abstract question. This additional context helps the LLM understand the alignment between values and schema elements without exposing any concrete tokens. 

A key point is that, once the schema linking map is generated, abstraction reduces to simple text substitution. This method preserves the reference information in the question while allowing for accurate abstraction inversion of an LLM's output.




\subsection{SQL Generation}
Using the abstracted question $\ques'$ and database schema $\schema'$ produced in the previous steps, a remotely hosted LLM is prompted to generate the corresponding abstracted SQL query $\sql'$. The generation prompt used in the experiments is provided in Appendix \ref{app:sql-prompt}. This prompt contains only $\ques'$ and $\schema'$, with no additional sensitive information. Since the LLM only sees abstract symbols, the generated SQL query is also expected to follow the abstract form. An example of such an abstract SQL query is shown in Figure \ref{fig:pipe}.
To address minor errors in the generated abstract SQL, a self-correction mechanism, commonly used in \ts\ pipelines, is employed \citep{din}. In this step, the LLM is prompted with $\ques'$, $\schema'$, and $\sql'$ and is instructed to identify and correct any potential issues. The prompt used for this step is included in Appendix \ref{app:repair-prompt-abs}.

\subsection{SQL Reconstruction}
In the final stage, the abstract SQL query $\sql'$ is mapped back to its concrete form $\sql$. Using the symbol lookup table created during abstraction, all abstract symbols are replaced with their corresponding concrete values. The result is an executable SQL query free of abstract identifiers. To further improve accuracy, an additional self-correction step is applied using a local SLM. The concretized SQL query is executed on the target database, and its execution result, along with $\ques$, $\schema$, and the concretized query $\sql$, is passed to the SLM to correct any remaining errors. The full prompt for this step is provided in Appendix \ref{app:repair-prompt-conc}.
\section{Experiments}
\label{sec:experiments}

In this section, experimental results are reported comparing \name\ with several state-of-the-art \ts\ frameworks. We use the \bird~dataset, a widely used benchmark consisting of NL questions paired with ground-truth SQL queries, along with the corresponding databases that enable execution-based evaluation of generated queries \citep{bird}. To demonstrate the gap between state-of-the-art LLM- and SLM-based \ts\ models on more complex queries, a challenging subset of $300$ entries is selected from the \bird~development split. In particular, we selected examples that involve complex SQL patterns, such as nested queries, set operations (e.g., INTERSECT), and multiple joins. To generate ground-truth data for privacy measurements, GPT-4.1 \citep{gpt4.1} is used to annotate tokens that should be abstracted, followed by human review and correction. For \name, Qwen-2.5-7B-Instruct \citep{qwen2.5} serves as the trusted local SLM, and GPT-4.1 is used as the untrusted LLM for generating SQL queries from abstracted prompts.

We evaluate \name\ under two privacy policies.

\begin{itemize}
\setlength{\leftskip}{\itemizelength}
    \item $\policy_F$: The full privacy policy, defined in Section \ref{sec:prob}, where the entire database schema and associated values are considered sensitive.
    \item $\policy_C$: A category-based policy where only tokens related to concepts of person names, occupations, and locations are abstracted. The formal definition is provided in Appendix \ref{sec:category-def}.
\end{itemize}

\subsection{Baselines}
We compare \name\ against several baselines, including three state-of-the-art \ts\ frameworks and direct LM prompting.  The direct prompting baseline applies a simple few-shot prompt that asks the LM to generate a SQL query based on the given NL question and database schema. \dail\ \citep{dail}, which holds the top position on the Spider benchmark \citep{spider} for open-source solutions at the time of writing, builds on few-shot prompting with additional strategies to improve SQL generation accuracy. \din\ \citep{din}, which holds the second-best ranking among open-source frameworks, decomposes \ts\ translation into smaller tasks to improve the accuracy. For direct prompting, \dail, and \din, we consider both a trusted setting, where Qwen2.5-7B-Instruct is used as the backbone LM and an untrusted setting, where GPT-4.1 is used. 
Finally, MSc-SQL \citep{msc} is designed specifically for SLM-only settings. It samples a few candidate SQL queries from different models and then selects the best candidate by prompting additional SLMs. For experiments leveraging MSc-SQL, we use the fine-tuned SLMs released by the authors based on Gemma-2-9B-it \citep{gemma}, Llama-3-8B \citep{llama}, and Mistral-7B-Instruct-v0.2 \citep{mistral}.

\subsection{Metrics}
We evaluate each baseline on utility, efficiency, and privacy metrics. For utility evaluation, we use execution accuracy, as defined in the \bird~benchmark. Efficiency is measured by average token usage per query generation. For privacy, we define two metrics.

Masking Recall (MR): This metric is defined as the ratio of correctly abstracted tokens to the total number of ground-truth sensitive tokens in the NL question $\ques$. Higher values indicate more protection of the sensitive tokens, resulting in better privacy.

Re-identification Score (RI): This metric captures the proportion of abstracted tokens in the NL question $\ques'$ that cannot be re-identified by an adversary. Specifically, we prompt GPT-4.1 with the abstracted NL question $\ques'$ and schema $\schema'$ and instruct it to infer the original tokens. The score is then computed as the ratio of the tokens that cannot be recovered by the LLM to the total number of abstract tokens in $\ques'$. 

Formal definitions of these metrics are provided in Appendix \ref{sec:priv-metrics}.

\begin{table} [t]
\centering
\captionsetup{skip=10pt}
\begin{tabular}{lccc}
\toprule
Framework & Execution Accuracy & Token Usage & Trusted \\
\midrule
Direct Prompting + \qwen                                  & 34.33\% & 1,380 & Yes \\
\dail~+ \qwen                                   & 44.33\% & 3,492 & Yes \\
Fine-Tuned MSc-SQL                      & 48.33\% & 8,342 & Yes \\
\din~+ \qwen                                   & 50.66\% & 24,812 & Yes \\
\textbf{\name~($\policy_F$)}             & \textbf{55.66\%} & \textbf{6,114}  & \textbf{Yes} \\
\textbf{\name~($\policy_C$)}   & \textbf{62.66\%} & \textbf{6,757}  & \textbf{Yes} \\
\midrule
\dail~+ \gpt                                  & 63.33\% & 3,385 & No \\
\din~+ \gpt                                    & 73.66\% & 23,036 & No \\
Direct Prompting + \gpt                                    & 75.66\% & 1,352 & No \\
\bottomrule
\end{tabular}
\caption{Execution accuracy and token usage of \name\ compared to other \ts\ frameworks on a subset of the \bird~development set. \name\ outperforms SLM-based frameworks and achieves better token efficiency than MSc-SQL and \din. By adopting a more permissive privacy policy, \name~($\policy_C$) improves execution accuracy while still preserving privacy. Pure LLM-based methods have the best accuracy, but do not protect sensitive data.}

\label{table:acc&tok}
\end{table}

\subsection{Results}
\signpost{Accuracy} Table \ref{table:acc&tok} shows the execution accuracy across baselines. Untrusted LLM-based approaches perform significantly better than trusted SLM-based methods, which is expected given the 
additional capacity of LLMs. However, the direct use of LLMs exposes sensitive data to privacy risks. 
\name\ outperforms all trusted SLM-only approaches, achieving an execution accuracy of 55.66\% under the full privacy policy, $\policy_F$, which is $5$ percentage points higher than the 
next best trusted baseline, \din. This improvement is primarily because \name\ leverages LLMs in a privacy-preserving manner. However, there remains a $\sim$20 point gap between \name\ and the best-performing untrusted baseline. We also observe in the experiments that the direct prompting GPT-4.1 outperforms both \din\ and \dail. This is a surprising result, as these methods are supposed to enhance simple LLM prompting and yield higher accuracy. We discuss this further in Appendix \ref{sec:results-disc}.

\signpost{Efficiency} 
Table \ref{table:acc&tok} also reports the average token usage of each framework per SQL generation.
For frameworks that make more than one LM call, we compute the total token usage across all calls.
As shown in the table, direct prompting with Qwen2.5 and GPT-4.1 exhibits the lowest token usage, followed by \dail\ and MSc-SQL, while \din\ consumes the most. \name\ uses a competitive number of tokens compared to SLM-based frameworks, especially when considering execution accuracy gains. 
For instance, it requires a quarter of the tokens consumed by the next best trusted model. The additional token usage in \name\ primarily comes from the intermediate linking steps and error corrections. 

\begin{figure}[t]
    \hspace{0.75cm} 
    \includegraphics[width=0.8\linewidth]{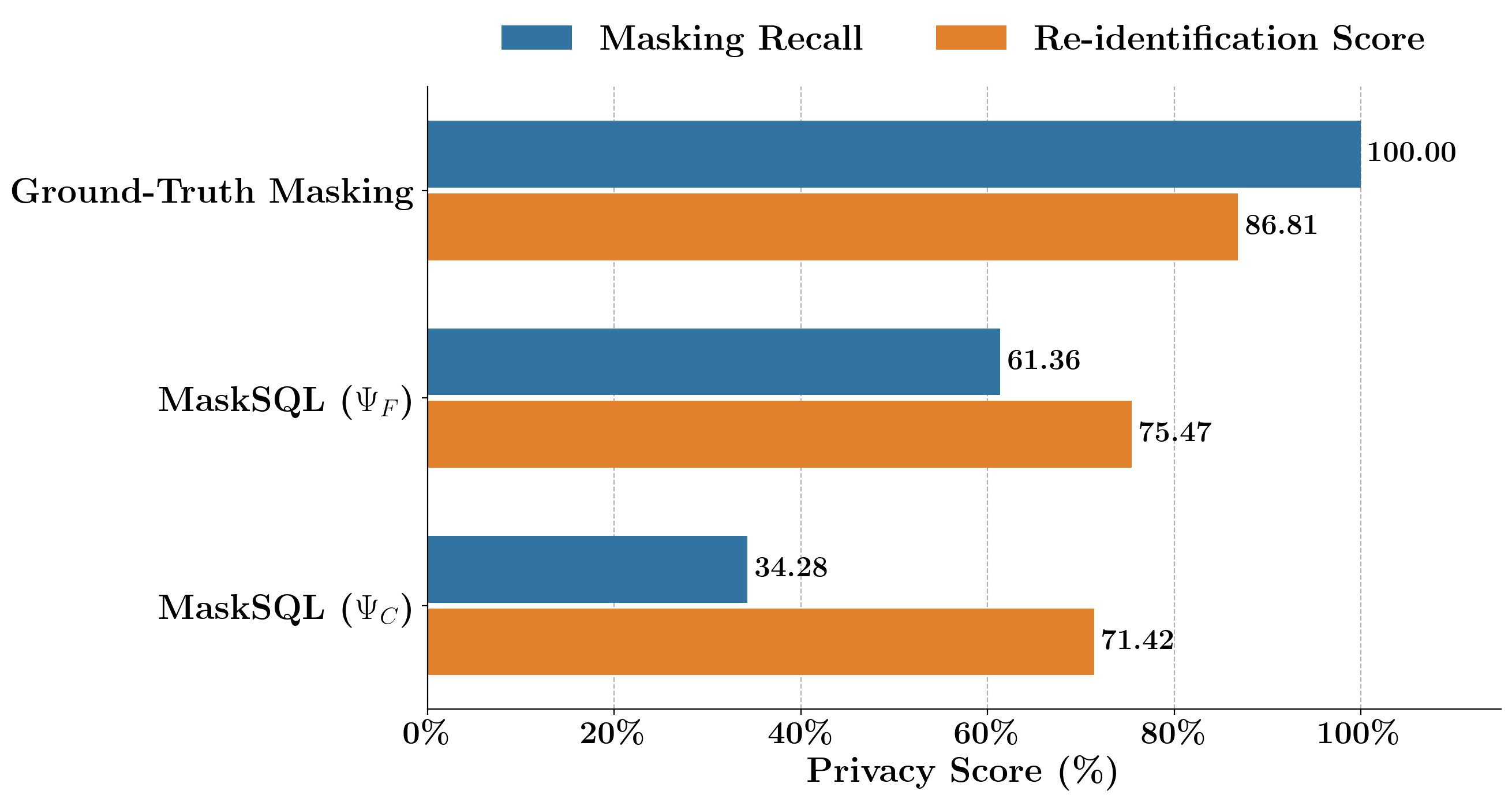}
    \caption{Privacy metrics of \name\ compared to ground-truth masking. 
    Higher values indicate stronger privacy preservation.}
    \label{fig:priv}
\end{figure}

\signpost{Privacy}
Figure \ref{fig:priv} presents the privacy scores of \name\ measured by the masking recall and re-identification score metrics.
In Figure \ref{fig:priv}, \textit{ground-truth masking} refers to abstracting all ground-truth tokens, providing an upper bound for masking recall and re-identification scores.
Under the full policy, $\policy_F$, \name\ achieves a masking recall of 61.36\%, demonstrating that a large portion of sensitive tokens are effectively abstracted. 
The re-identification score measures robustness against adversarial recovery of the abstract tokens. With GPT-4.1 acting as the attacker, \name\ achieves a re-identification score of 75.47\%, indicating that three-quarters of abstracted tokens could not be inferred from context. With ground-truth masking, this score is $\sim$86\%. The relatively small gap highlights that \name\ is robust to contextual information leakage despite not abstracting all tokens.

We further analyze the impact of using a more permissive privacy policy. As shown in Table \ref{table:acc&tok}, using the category-based policy, $\policy_C$, increases execution accuracy by $7$ points compared to $\policy_F$ at the cost of a $27$ point drop in masking recall, which is expected, since fewer tokens are abstracted when only tokens related to specific concepts (name, location, and occupation) are considered. This highlights the flexibility of policy-based abstraction in trading off privacy for higher accuracy, depending on the user’s needs. Note that for \name\ under the $\policy_C$ setup, masking recall is computed as the ratio of correctly abstracted tokens in $\policy_C$ to all ground-truth sensitive tokens, rather than restricting the ground-truth tokens to only those included in the policy. Additionally, the re-identification score decreases by only $4$ points, indicating that while fewer tokens are abstracted overall, the remaining abstracted tokens remain difficult for the adversary to recover.

\section{Conclusion}
In this work, we introduced \name, a framework for \ts\ translation that employs an abstraction-based privacy mechanism to preserve sensitive information according to user-defined policies when interacting with LLMs. 
Rather than relying solely on either SLMs or LLMs, \name\ follows a hybrid design that combines private local processing with SLMs and leverages the reasoning ability of LLMs.
By supporting flexible privacy policies, \name\ also enables a privacy-utility trade-off that allows users to tailor abstraction according to their needs. The experiments show that \name\ outperforms state-of-the-art SLM-based methods in terms of execution accuracy. Additionally, we implemented a category-based privacy policy that considers only sensitive tokens associated with specific semantic categories to be abstracted. The experimental results demonstrate that this more lenient policy achieves higher execution accuracy than the full privacy policy, highlighting the potential of controlled trade-offs between privacy and utility.


There are several promising directions for future work to be explored. 
First, since only a subset of tokens is abstracted within the NL question, the surrounding context may be exploited to re-identify the abstract tokens. Currently, we quantify this privacy risk by measuring the re-identification score. However, additional privacy measures can be applied to mitigate such potential leakage.
Next, while \name\ uses an LLM backbone for SQL generation, its accuracy still falls behind other LLM-based frameworks. Future work will explore fine-tuning SLMs, improved schema linking, and token-efficient strategies to reduce this gap.
Further, we plan to consider integrating provable privacy mechanisms, such as differential privacy, into the \name\ pipeline. In addition to the privacy measurements leveraged in this work, provable guarantees enhance the user's trust in the system.
Finally, the abstraction-based approach in \name\ is not limited to the \ts\ translation task. We envision the abstraction approach to be applicable to tasks such as code generation, debugging, and data analysis, where reference and relational information are sufficient to perform the task. Characterizing the class of tasks to which this approach applies will be the subject of future work.





\bibliographystyle{plainnat}
\bibliography{refs}

\appendix

\section{Category-Based Policy Definition}
\label{sec:category-def}
Let $\policy_{F}$ be the full policy under which the entire schema and database values are abstracted.
Let $\mathcal{L}$ be the set of semantic categories. Each category in $\mathcal{L}$ is a label for a certain concept, such as ``person's name'' or ``locations''. We define the category-based policy $\policy_{C}$ for categories $\mathcal{L}$ as
\begin{align*}
    \policy_{C} = \{w \in \policy_{F} \mid l(w) \in \mathcal{L} \},
\end{align*}

where $l: \mathbb{W} \rightarrow \mathcal{L} \cup \{\varepsilon \}$ is a labeling function that assigns a category from the set $\mathcal{L}$ to a token $w \in \mathbb{W}$. In the case that a token does not match any category, it is mapped to $\varepsilon$. We assume that the labeling function is implemented with semantic text classification \cite{classif}. Here, we use Qwen2.5-7B to classify the tokens into categories based on their semantics.

\section{Privacy Metrics Definitions}
\label{sec:priv-metrics}

\subsection{Re-identification score (RI)}
This metric is calculated by simulating an attack on the masked question.
More specifically, an LLM is prompted with the masked NL question and database schema and instructed to infer the masked tokens.
Then, the percentage of the masked tokens that could not be re-identified is calculated as the score. Formally, let $\mathbb{W}$ be the set of all natural language tokens and $\mathbb{S} \subseteq \mathbb{W}$ be the set of masked symbols.
Let $\ques = w_1,w_2,\ldots,w_n$ be an NL question where $w_i \in \mathbb{W}$ for $1\leq i \leq n$.
We define $\ques_s = \{ w_i \mid 1 \leq i \leq n \}$ as the set of tokens used in $\ques$.
We assume that questions do not include masked tokens, i.e., $\ques_s \cap \mathbb{S} = \emptyset$.
For a database schema $\mathcal{S}$, we define masking functions as $f: \mathbb{W} \rightarrow \mathbb{W}$, which maps each token of a question to either a masked symbol or the token itself. Let $f^{-1}$ denote the inverse of $f$, which returns the original token for a masked token. For a question $\ques$, let $f(\ques) = f(w_1), f(w_2), \ldots, f(w_n)$ be its masked version.
For a question $\ques$ and masking function $f$, let $\ques' = f(\ques)$ be its masked version, and $G$ be the re-identified question generated by a simulated attack on $\ques'$. Let $\ques_s'$ and $G_s$ be the set of tokens for $\ques'$ and $G$, respectively. 
We define the re-identification score as follows:
\begin{align}
    R_{f}(\ques_s', G_s) & = 1 - \frac{\lvert\{ w \in \ques_s \mid f(w) \in \mathbb{S}\} \cap G_s \rvert}{\lvert \ques_s' \cap \mathbb{S} \rvert}     
\end{align}
where the set builder notation is used to denote the set of original tokens in the $\ques_s$ that have been masked by $f$, and $\ques'_s \cap \mathbb{S}$ is the set of all masked tokens in $\ques_s'$.
A score of 1 means none of the masked tokens could be inferred by the attacker. Here, we use exact string matching and not semantic or distance-based methods.

We calculate the re-identification score for each masked question and report the average over all questions as the score for the whole dataset.


\subsection{Masking Recall (MR)}
Masking recall is defined as the ratio of the masked tokens to the total number of tokens in the ground-truth masking.
Formally, let $f,f_g$ be two masking functions where $f_g$ is the ground-truth, i.e., a mapping that masks all ground-truth sensitive tokens.
For a question $\ques$ and its set of tokens $\ques_s$, we define the masking recall as follows:
\begin{align}
    M_{f}(\ques_s) & = \frac{
    \{ w \in \ques_s \mid f_g(w) \in \mathbb{S} \wedge f(w) \in \mathbb{S} \}
    }{\lvert f_g(\ques_s) \rvert}
\end{align}

A score of 1 indicates that the tool successfully masked all required tokens. Similarly, we calculate the average of the scores to report it for a dataset.

\section{Ablation Study}
\label{sec:abl}

To quantify the contribution of each component in the \name~pipeline, we conducted an ablation study by removing each stage and measuring how accuracy and privacy scores are changed. The results, presented in Table \ref{table:abl}, demonstrate that each component plays a crucial role in the \name\ pipeline (Figure \ref{fig:pipe}).

\begin{table}[H]
\centering
\captionsetup{skip=10pt}
\begin{tabular}{lccc}
\toprule
Stage & EX & RI & MR \\
\midrule
Complete Pipeline         & \textbf{55.66\%} & 75.47\% & 61.36\% \\
Schema Filtering          & 53.66\% & 76.49\% & 58.56\% \\ 
Value Detection           & 53.33\% & 82.17\% & 64.45\% \\ 
Value Linking             & 52.33\% & 74.97\% & 48.64\% \\ 
Abstraction               & 47.00\% & \textbf{86.79}\%  & \textbf{65.45}\%  \\ 
LLM Correction            & 51.33\% & 75.47\% & 61.36\% \\ 
SQL Reconstruction        & 34.33\% & 75.47\% & 61.36\% \\ 
SLM Correction            & 35.33\% & 75.47\% & 61.36\% \\ 
\bottomrule
\end{tabular}
\caption{Ablation study results showing the effect of removing each component of the \name\ pipeline on the final execution accuracy (EX), re-identification score (RI), and masking recall (MR).}
\label{table:abl}
\end{table}

\signpost{Schema Filtering}
For this experiment, we remove the schema filtering at the beginning of the pipeline. As a result, the whole database schema is always passed to the further stages. As shown in the table, removing schema linking step decreases the execution accuracy and masking recall.

\signpost{Value Detection}
Here, we remove the separate stage for detecting the values in the question.
Removing this stage slightly decreases the execution accuracy.

\signpost{Value Linking}
In this experiment, we remove the separate stage for linking the values of the question. While value linking slightly decreases the execution accuracy, it has a significant negative impact on the masking recall and decreases it by
$\sim 13\%$.

\signpost{Abstraction}
In the \name\ pipeline, we use separate stages for schema linking and abstraction.
To measure the effect of this separation, we removed the schema linking stage and used an SLM to abstract away the question with respect to the database schema.
While abstraction with SLM and without a separate schema linking step improves the privacy measures, it decreases the execution accuracy by 
$\sim 8\%$.

\signpost{LLM Correction}
By removing the abstract SQL correction stage, we see a slight decrease in execution accuracy.

\signpost{SQL Reconstruction}
Similar to abstraction, to measure the effect of the deterministic unmasking process, we prompt SLM to unmask the SQL query and also correct the errors in the query if needed. As shown in the table, replacing the deterministic unmasking stage with language model prompting significantly reduces the execution accuracy by $\sim 21\%$.

\signpost{SLM correction}
The final correction stage prompts the SLM to fix errors in the reconstructed SQL query, and, as shown in the table, it has a significant impact on the execution accuracy.
Removing this correction step causes a $\sim 20\%$ drop in execution accuracy.

\section{Discussion of Experimental Results}
\label{sec:results-disc}
Table \ref{table:acc&tok} shows the execution accuracy of different frameworks. As shown in the figure, direct prompting to GPT-4.1 outperforms all evaluated models. This is a surprising result, as both \din\ and \dail\ are also utilizing GPT-4.1 as the backend LLM and performing additional processing and improvements to enhance accuracy.
The direct method uses the prompt shown in Figure \ref{app:sql-prompt}, but with the concrete examples instead of the abstract ones shown in the figure.
Note that the BIRD benchmark website reports \din and \dail submissions based on GPT-4, rather than the later version GPT-4.1. Moreover, our experiments use a different evaluation set. These differences explain why GPT-4.1 surpassed them in our results, even though they achieved top ranks on the official BIRD leaderboard.

\section{Additional Figures}
\label{sec:extra-figs}

\begin{figure}[H]
\begin{minted}[fontsize=\footnotesize, breaklines, breakanywhere]{sql}
-- NL Question:
-- List the power of superheroes with height greater than 80% of the average height of all superheroes. 
-- Power of superheroes refers to power_name;
-- Height greater than 80% of the average height of all superheroes = height_cm > MULTIPLY(AVG(height_cm), 0.8);

    
-- Qwen2.5 Output (Incorrect):
SELECT 
    [superpower].[power_name]
FROM 
    [hero_power]
JOIN 
    [superpower] ON [hero_power].[power_id] = [superpower].[id]
JOIN 
    [superhero] ON [hero_power].[hero_id] = [superhero].[id]
GROUP BY 
    [superpower].[power_name]
HAVING 
    [superhero].[height_cm] > AVG([superhero].[height_cm]) * 0.8;
        
-- GPT-4.1 Output (Correct):
SELECT 
    [superpower].[power_name]
FROM 
    [hero_power]
JOIN 
    [superhero] ON [hero_power].[hero_id] = [superhero].[id]
JOIN 
    [superpower] ON [hero_power].[power_id] = [superpower].[id]
WHERE 
    [superhero].[height_cm] > (
        0.8 * (SELECT AVG([superhero].[height_cm]) FROM [superhero])
    );
\end{minted}

\caption{An example of an SQL query that requires advanced constructs like nested queries, which Qwen2.5-7B failed to handle properly. This example is extracted from our experiments.}
\label{fig:hard-sql-example}
\end{figure}

\begin{figure}[ht]
\captionsetup[subfigure]{labelformat=empty}
\begin{subfigure}[t]{0.48\textwidth}
\caption{\textbf{NL Question ($\ques$)}}
\begin{minted}[fontsize=\footnotesize, breaklines, breakanywhere]{text}
How many patients did the New York Hospital admit with HIV status as positive?


\end{minted}
\end{subfigure}
\begin{subfigure}[t]{0.48\textwidth}
\caption{\textbf{Abstract NL Question ($\ques'$)}}
\begin{minted}[fontsize=\footnotesize, breaklines, breakanywhere]{text}
How many T1 did the V1 T3 with C3 as V2?; V1 is a value of the column C7; V2 is a value of the column C3


\end{minted}
\end{subfigure}
\begin{subfigure}[t]{0.48\textwidth}
\caption{\textbf{SQL Query ($\sql$)}}
\begin{minted}[fontsize=\footnotesize, breaklines, breakanywhere]{sql}
SELECT count(P.pid) 
FROM Patient AS P
JOIN Admission AS A ON P.pid = A.pid
JOIN Hospital AS H ON A.hid = H.hid
WHERE H.name = "New York Hospital"
AND P.hiv_status = 1


\end{minted}
\end{subfigure}
\begin{subfigure}[t]{0.48\textwidth}
\caption{\textbf{Abstract SQL Query ($\sql'$)}}
\centering
\begin{minted}[fontsize=\footnotesize, breaklines, breakanywhere]{sql}        
SELECT count(T1.C1) 
FROM T1 
JOIN T3 ON T1.C1 = T3.C10
JOIN T2 ON T3.C11 = T2.C6
WHERE  T2.C7 = 'V1'
AND T1.C3 = 'V2';


\end{minted}
\end{subfigure}
\begin{subfigure}[b]{0.45\textwidth}
\caption{\textbf{Database Schema ($\schema$)}}
\begin{minted}[fontsize=\footnotesize]{yaml}        
'Patients':
    'pid':
        primary_key: true
        type: integer
    'name': text
    'hiv_status': integer
    'diagnosis': text
    'treatment': text
'Hospital':
    'hid':
        primary_key: true
        type: integer
    'name': text
    'address': text
'Admissions':
    'aid':
        primary_key: true
        type: integer
    'pid':
        foreign_key: 'Patients.pid'
        type: integer
    'hid':
        foreign_key: 'Hospital.hid'
        type: integer
    'date': date
    
\end{minted}
\end{subfigure}
\begin{subfigure}[b]{0.45\textwidth}
\caption{\textbf{Abstract Database Schema ($\schema'$)}}
\begin{minted}[fontsize=\footnotesize]{yaml}        
'T1':
    'C1':
        primary_key: true
        type: integer
    'C2': text
    'C3': integer
    'C4': text
    'C5': text
'T2':
    'C6':
        primary_key: true
        type: integer
    'C7': text
    'C8': text
'T3':
    'C9':
        primary_key: true
        type: integer
    'C10':
        foreign_key: 'T1.C1'
        type: integer
    'C11':
        foreign_key: 'T2.C6'
        type: integer
    'C12': date
    
        \end{minted}
    \end{subfigure}
    \caption{Abstract question, database schema, SQL query for Example \ref{running_example}.}
    \label{fig:running-example-abs}
\end{figure}

\section{Prompt Templates}
\label{sec:prompts}

\subsection{SQL Generation Prompt}
\label{app:sql-prompt}
\begin{Verbatim}[breaklines]
You are an SQL generation assistant. Given 

(1) NL Question: a natural-language question about a dataset and 

(2) DB Schema: the database’s schema expressed in YAML

produce a single SQL SELECT statement that answers the question.

Input Format:
- DB Schema: given in YAML format, where top-level keys are table names; each table lists its columns and their data types.
- Column names are case-sensitive exactly as shown in the schema.
- Each column might be a primary key or a foreign key.
- For foreign key columns, the fully qualified name of the referenced column is given.

Output Rules:
- Table and column names specified in the database schema are already wrapped in brackets. You should use them with the brackets.
You should not remove the brackets when using them in the SQL.
- Each reference to a table or column name should be of the form [table_name] or [table_name].[column_name].
- Output ONLY the SQL query (no extra explanation or text).
- Use fully qualified column names: table.column.
- Only reference tables/columns that exist in the provided schema.
- Do not include any comments.
- For column names with spaces, wrap them in backticks, e.g., "WHERE `car model` = 'bar'" instead of "WHERE car model = 'bar'".

Here are some examples:

...
-----------------------------------
#### Example 2

**NL Question:**
Among the [V1] [T1], how many of them have a [C2] of [V2]? [V1] refers to [C2] = [V1].

**DB Schema:**
[T1]:
    [C1]: text
    [C2]: real
    [C3]:
        primary_key: true
        type: integer

**Output:**
`SELECT COUNT(*) FROM [T1] WHERE [T1].[C1] = [V1] AND [C2] = [V2]

-----------------------------------
...

Now, generate an SQL query for the following question and database schema:
Inputs:
NL Question: {NL_QUESTION}
DB Schema: {DB_SCHEMA}
\end{Verbatim}

\subsection{Self-Correction Prompt (Abstract)}
\label{app:repair-prompt-abs}
\begin{Verbatim}[breaklines]
You are an SQL database expert tasked with debugging an SQL query. 
A previous attempt to predict an SQL query given a masked NL question and DB schema did not yield the correct results in some cases. 
Either due to errors in execution or because the result returned was empty or unexpected. 
Your task is to analyze the masked SQL query given the corresponding database schema and the NL question. 
and fix any errors in the query if they exist.
You should then provide a corrected version of the SQL query.
Note that there may be errors in how the NL question tokens were linked and masked with the database schema elements.
As a result, the masked SQL query might contain inaccuracies based on these incorrect mappings, 
and part of your task is to consider these issues as well.

**Procedure:** 
1. Review Database Schema: 
    - Examine the database schema to understand the database structure.
    - Database schema is given in YAML format, where top-level keys are table names; each table lists its columns and their data types.
    - Each column might be a primary key or a foreign key.
2. Analyze Query Requirements: 
    - NL Question: Consider what information the query is supposed to retrieve. 
    - Predicted SQL Query: Review the SQL query that was previously predicted and might have led to an error or incorrect result. 
3. Correct the Query: 
    - Modify the SQL query to address the identified issues, ensuring it correctly fetches the requested data according to the database schema and query requirements.
    
**Output Format:** 
Present your corrected query as a single line of SQL code. 
Ensure there are no line breaks within the query.
Do not include any explanations, comments, or extra text.

Here are some examples: 

-------------------------------
Example 1:

NL Question:
Among the [V1] [T1], how many of them have [C2] of zero? [V1] is a nationality of [C1] = [V1];

Database Schema:
'[T1]':
    '[C1]': text
    '[C2]': real
    '[C3]':
        primary_key: true
        type: integer

The predicted SQL query was: 
SELECT COUNT(*) FROM [T1] WHERE [T1].[C1] = '[V1]' AND [C2] = 0

The corrected SQL query is:
SELECT COUNT(*) FROM [T1] WHERE [T1].[C1] = '[V1]' AND [C2] = 0
-------------------------------

======= Your task ======= 
************************** 
Database schema:
{schema} 
************************** 
The original question is: 
NL Question: {question} 
The predicted SQL query: {sql} 
************************** 
Based on the NL question, database schema, and the previously predicted SQL query, 
Analyze the query and question, and fix the SQL query if needed.

\end{Verbatim}

\subsection{Self-Correction Prompt (Concrete)}
\label{app:repair-prompt-conc}
\begin{Verbatim}[breaklines]
You are an SQL database expert tasked with correcting an SQL query. 
A previous attempt to run a query did not yield the correct results, 
either due to errors in execution or because the result returned was empty or unexpected. 
Your role is to analyze the error based on the provided database schema and the details 
of the failed execution, and then provide a corrected version of the SQL query.

**Procedure:** 
1. Review Database Schema: 
    - Examine the database schema to understand the database structure.
    - Iterate through each column and table name in the schema to make sure that it is correct.
    - Some table or column names may have white space; you should not change these and use a different name.
    - Database schema is given in YAML format, where top-level keys are table names; each table lists its columns and their data types.
    - Column names are case-sensitive, exactly as shown in the schema.
    - Each column might be a primary key or a foreign key.
    - For foreign key columns, the fully qualified name of the referenced column is given.
2. Analyze Query Requirements: 
    - Original Question: Consider what information the query is supposed to retrieve. 
    - Executed SQL Query: Review the SQL query that was previously executed and led to an error or incorrect result. 
    - Execution Result: Analyze the outcome of the executed query to identify why it failed (e.g., syntax errors, incorrect column references, logical mistakes). 
3. Correct the Query: 
    - Modify the SQL query to address the identified issues, ensuring it correctly fetches the requested data according to the database schema and query requirements.
    - For column names with spaces, wrap them in backticks, e.g., "WHERE `car model` = 'bar'" instead of "WHERE car model = 'bar'".

**Output Format:** 
Present your corrected query as a single line of SQL code. 
Ensure there are no line breaks within the query.
Do not include any explanations, comments, or extra text.

Here are some examples: 

-------------------------------
Example 1:
Question:
Among the German customers, how many of them have a credit limit of zero? German is a nationality of the country = 'Germany'; CREDITLIMIT = 0.

Database Schema:
'[customers]':
    '[country]': text
    '[creditlimit]': real
    '[customernumber]':
        primary_key: true
        type: integer

The SQL query executed was: 
SELECT COUNT(*) FROM [customers] WHERE [customers].[country] = 'German' AND [creditlimit] = 0

Output:
SELECT COUNT(*) FROM [customers] WHERE [customers].[country] = 'Germany' AND [creditlimit] = 0

The execution result: 
[]
-------------------------------

======= Your task ======= 
************************** 
Database schema:
{schema} 
************************** 
The original question is: 
Question: {question} 
The SQL query executed was: {sql} 
The execution result: {exec_res} 
************************** 
Based on the question, table schema, and the previous query, analyze the result and try to fix the query.

\end{Verbatim}

\end{document}